# Development of High-Quality Polymer/Air-Bragg Micromirror Structures for Nanophotonic Applications

Chirag Chandrakant Palekar[1*], Manan Shah[2], Stephan Reitzenstein[1], Arash Rahimi-Iman[2*]

[1]Institute of Solid State Physics, Technische Universität Berlin, D-10623, Germany

[2]I. Physikalisches Institut and Center for Materials Research, Justus-Liebig-Universität Gießen, D-35392, Germany.

*Emails: c.palekar@tu-berlin.de and arash.rahimi-iman@physik.jlug.de*

**We report the design, nanofabrication, and characterization of high-quality polymer-based micromirror structures employing the 3D two-photon polymerization lithography technique. Compared to conventional microcavity approaches, our innovative concept provides microstructures which allow fast prototyping. Moreover, our polymer-based mirrors are cost effective, environmentally sensitive, as well as compatible with a wide range of wavelengths from NIR to the telecom C-band. We demonstrate polymer/air distributed Bragg reflectors and full microcavity structures with up to 14 mirror pairs with a target wavelength of 1550 nm and a reflectivity close to 99%. Additionally, our 3D printed micromirrors are reproducible, mechanically stable, and enable hybrid nanophotonic devices based on quantum dots, molecules or 2D quantum materials as active medium.**

Light confinement, reflection and transmission are typically at the heart of many photonic experiments and nanophotonic device functionalities. Typically, high-quality mirrors with design-wavelength reflectivity beyond those of element metals such as gold, silver or aluminum are obtained using concepts from wave physics for tailored interference effects [1,2]. Commonly, 1D or 2D photonic crystals, i.e. vertical Bragg gratings or in-plane periodic hole arrangements [3], respectively, are used to obtain photonic stop bands and light-field control with the desired directionality, e.g. for light-matter interaction (LMI) scenarios [4,5] in emitter-cavity systems.

Various types of photonic microcavities can provide high-quality resonator modes and strongly-confined light fields, which can be used in nonlinear optics [6,7] or nanolaser development [8,9]. They are also interesting for quantum nanophotonics [10] and fundamental cavity quantum electrodynamics (cQED) experiments [11,12–15]. Modern concepts such as efficient quantum light sources and quantum memories [16,17], exciton-polariton condensate devices [4,18], polariton chemistry [19] as well as photonic quantum simulators [20] and topological photonics [21–23] are hardly achievable without suitable microcavities.

Typically, the desired optical properties of microcavities are obtained using dielectric and semiconductor distributed Bragg mirrors (DBRs), which can be controllably grown in the vertical direction by established layer-deposition techniques and crystal epitaxy, respectively, leading to high-reflectivity photonic stop bands and resonators with high Q-factors. More recently, highly versatile, spectrally tunable, and relatively simple open-cavity configurations have been used in a large spectrum of applications ranging from cQED to (fiber-coupled) optoelectronical or optomechanical devices [24–26] as well as sensors [27–29].

Here, we develop 3D printed polymer "air-Bragg" micromirrors and -cavities to complement the pool of structure options at the disposal of photonic experiments and light-field engineers. These flexible structures – both in terms of design and mechanically-induced property modifications – consist of alternating layers of air/vacuum and the photoresist polymer (resin), and are compatible with various substrates and different photonic materials to enhance light-matter interaction [30]. In addition, by employing an inexpensive polymer structure which might be used in a disposable fashion, one-time-use sensing devices or cavity-enhanced optical probes could be obtained for diagnostics based on polymer/air reflectors and polymer photonics. According to the interferometric working principle of the DBR, the achievable stopband reflectance depends on the refractive index contrast of the used materials. By employing polymer/air layer pairs in the DBR, one can obtain a relatively high refractive index modulation despite the rather low refractive index (approximately 1.5) of the polymer itself. Using 3D two-photon polymerization lithography (TPL) nanoprinting, we fabricate stable polymer-based DBRs featuring a high-reflectivity stopband in the near-infrared with both bridge- and coaxial-type structure designs on quartz substrates.

Remarkably, the TPL method has evolved into a reliable versatile platform for the creation of various (3D) microoptical components, even in the shape of conical microlenses [31], diffractive Fresnel lenses [32] or achromatic meta-fiber facets [33].

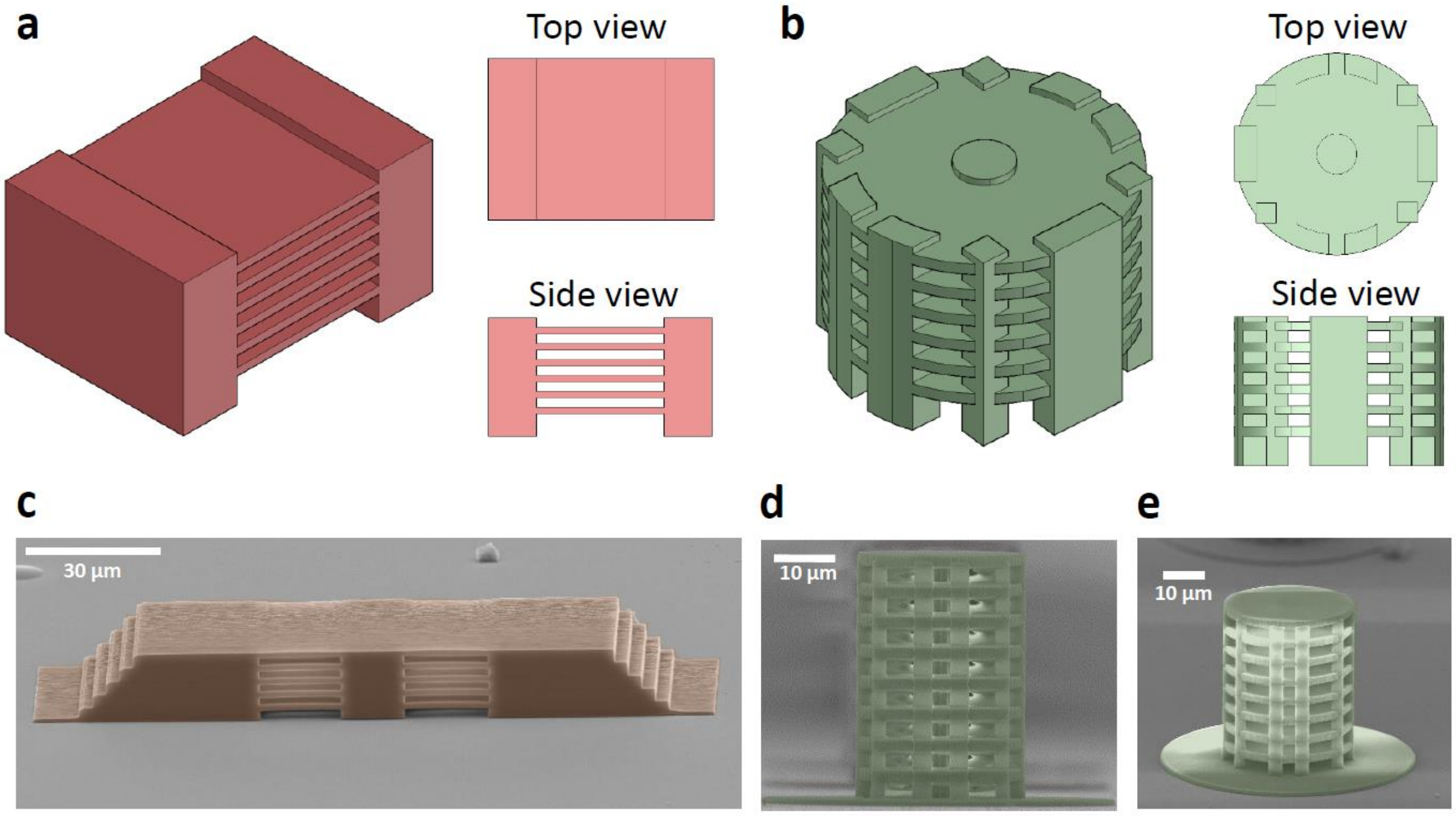

**Figure 1:** Schematic drawings of a (a) bridge- and (b) coaxial-type polymer/air-Bragg mirror and representative pseudo-color scanning-electron micrographs in (c-e). A two-photon lithography process defined the 3D microstructures with alternating multi-quarter-wavelength layers made of polymer and air/vacuum on quartz substrate.

In the proposed microresonator fabrication process, alternating layers are directly produced by 3D lithographically with the aid of a commercial TPL system (*Nanoscribe*), which provides high precision laser writing with feature resolution of the order of 100 nm and 3D voxel sizes as small as half a micrometer. For the design and fabrication of high-reflectivity air-Braggs in the spectral region of 900-950 nm and in the telecom C-band region of 1500-1550 nm, the refractive indices *n* = 1.51 and 1.56 [34], respectively, are considered for the polymer. Stacked air-polymer layer sequences are based on thickness calculations according to $d_{\mathrm{air/poly}} = i\,\lambda_0/(4\,n_{\mathrm{air/poly}})$, where $\lambda_0$ is the wavelength in vacuum, $n_{\mathrm{air/poly}}$ a layer's refractive index, and $i$ = 1,3,5…. In the following, $(\lambda_0/\,n_{\mathrm{poly}})$ is denoted $\lambda$ (for the high-index layers), whereas $n_{\mathrm{air}} = 1$ assumed ($\lambda_0$) (for the low-index layers).

**Figure 1** shows schematic drawings of the considered polymer/air-Bragg mirror designs, from which the (**a**) bridge-type model provides a rectangular reflector area, whereas the (**b**) coaxial-type model features a circular geometry. Below the pseudo-three-dimensional model images, three corresponding example scanning-electron-microscopy (SEM) images are displayed in (**c**-**e**) for the two design types. On the left, a doublet Bragg bridge structure (like in a linear photonic molecule) is depicted (**c**), whereas two different side views of a coaxial polymer/air-Bragg tower are shown on the right (**d, e**).

The bridge design features two rectangular support bars on both sides of the Bragg region with the "free-standing" polymer layers arranged in-between them to form the optically refractive region. On the other hand, in the coaxial design, co-centric disks of polymer (and "air") are inscribed into the resin on top of each other alongside the main center post and side support structures, which are intended for providing mechanical stability and rigidity. The support elements also act as polymer spacers, keeping the polymer layers apart from each other with certain air-gap thicknesses in accordance with our design parameters. Noteworthy, in future, the coaxial resonator design is intended to be also printed on fiber end facets, as the center pillar diameter can be scaled to match the fiber core diameter for the implementation of compact, plug-and-play optical devices, such as fiber-coupled single-photon sources [35].

For practical reasons, thin support structures are circularly arranged at the side of the coaxial polymer/air-Bragg towers as presented in **Fig. 1(b)** to prevent layer collapse and bunching, and to keep the air spacer intact. Their arrangement includes gaps to facilitate appropriate liquid exchange/extraction during the process of resin development and sample cleaning. Both aforementioned designs are fabricated to demonstrate the feasibility of polymer/air-Bragg mirrors as part of our efforts to achieve a good structural and optical performance, compatibility with different samples and reproducibility in terms of printing output. The linear bridge design was mainly used for reflectors with design wavelengths of around 900 nm, while the coax design was chosen for air-Braggs addressing fiber-based telecommunication applications.

We obtained design parameters and reflection spectra for the examined structures using the transfer matrix method (TMM)[27]. For the calculation of theoretical reflection spectra along with the standing wave patterns at the Bragg wavelength, polymer/air-Bragg reflectors were modeled as one-dimensional stack of alternating layers of characteristic thickness.

3D laser printing based on TPL allows one to laser-print the optical and photonic structures with nanoscale precision[36]. We employ *Nanoscribe*'s IP DIP resin for 3D TPL, in which a 63x microscope objective is immersed (NA = 1.4). The focus spot of the fs-laser scans the target region to inscribe user-defined geometries. Following TPL, the resin is developed with the commercially available *MR DEV 600* developer liquid (immersion bath) for 10 to 12 minutes and afterwards rinsed (in ambient air) with isopropanol for about 2 minutes. Gold sputtering by a plasma in vacuum is used to coat structures with an approximately 10-nm gold film for SEM studies, which allows resolving the printed features with sub-100 nm resolution.

**Figure 1(c)** shows a corresponding SEM image of a printed bridge-type Bragg reflector which has two similar neighboring Bragg regions – a doublet of reflectors for direct print-result comparison between two similar optical regions. In this example, side support constructs are laterally stepwise extended towards the substrate to improve the robustness of the structure in scenarios with mechanical pressure applied.

To analyze the fabricated structures and to model the expected reflectivity spectra using a TMM model, we extracted the thickness of the individual layers (cf. **Supplement Fig. S1**) for a specific reflector design from SEM images. Comparing the experimental measured spectral response of the structures with the theoretically predicted one allows deducing a generalized picture of the achievable structural and optical properties.

An ideal stack of alternating multi-quarter-wavelength air and polymer layers can result in the theoretically predicted spectral response displayed in **Fig. 2(a)** along with the corresponding standing-wave pattern plotted in **Fig. 2(b)** for a 5-mirror-pair structure with design wavelength of 935 nm.

Through numerical modeling, we compare air-Braggs at this near-infrared design wavelength featuring four different layer thicknesses. **Figure 2(a)** shows the expected reduction of calculated stopband width from roughly 100 nm to 30 nm with increasing layer thickness from 1 to 7 times $\lambda/4$, respectively. The modulation profile of the refractive index is indicated in **Fig. 2(b)** (right axis) together with the corresponding TMM-simulated resonant electric field distributions for the four thickness configurations. The electric field strength decreases exponentially when penetrating the Bragg region from the free-space side (2, 6, 10, 14 μm in **Fig. 2(b)** from top to bottom) to below 1/e strength (indicated by star symbol) inside the 3rd layer pair in all four cases. While the field amplitude is almost completely vanished after 5 pairs at position 0 (substrate interface), the absolute length of mode-penetration depth increases with increasing thickness parameter.

We chose the design parameters for our experimental reflectors considering the writing resolution of the used TPL system. Hence, the practical polymer/air-Bragg structures were printed with air and polymer thicknesses being of $7\lambda/4$ type, for both design wavelengths investigated below. As presented in Fig. 2(c) and Fig. 3(c), the calculated stopband width is about 25 nm and 50 nm at about 900 nm and 1550 nm, respectively.

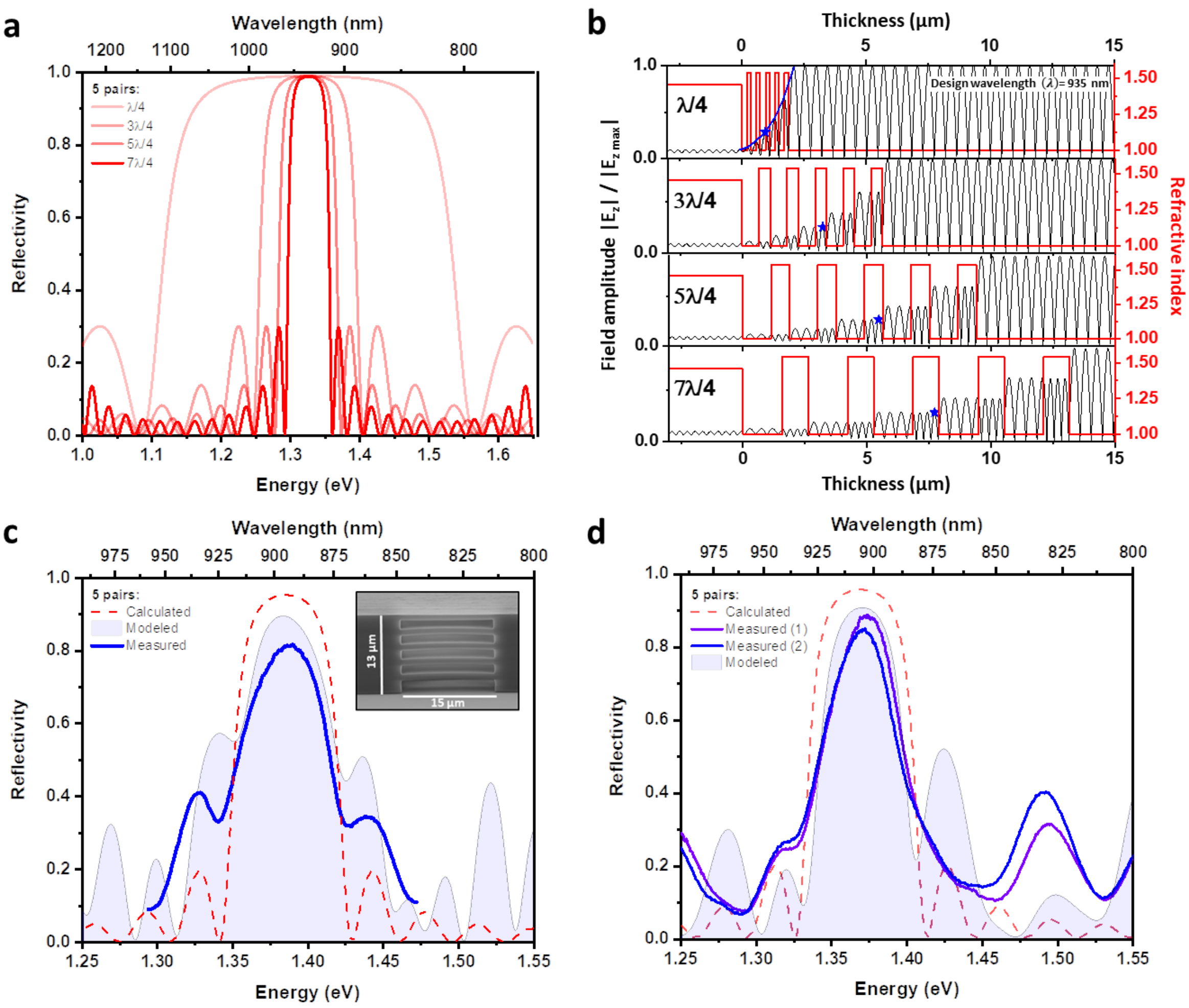

**Figure 2:** Theoretical reflectivity spectrum (a) and electric-field distribution (b) of 5-layer-pair Bragg mirror configurations with four different layer thicknesses, for a design wavelength of 935 nm. A black solid line indicates the field's exponential decay for the λ/4 polymer/air reflector, its drop below 1/e is marked by a star. (c) Comparison between measured, calculated and modeled reflection spectra, using printed Bragg layer thicknesses extracted from an SEM image of a sibling sample with the same layer design. The experimental reflectance data was obtained from nanoprinted air-Bragg structures on quartz substrate under white light illumination. Inset: Close-up SEM image of 5-pair bridge air-Bragg reflector with indicated structure size. (d) Reflectivity spectra measured for two print-improved neighboring Bragg structures, together with the theoretically predicted curve for an ideal 5 layer-pairs structure.

Next, we compare the optical performance of a printed polymer/air-Bragg reflector with the modeled reflectivity spectrum (shaded area curves) based on SEM-extracted thickness values of a nominally identical sibling sample. The experimental response of each reflector structure is obtained under white-light illumination in a micro-spectroscopy setup, employing a microscope objective with an NA of 0.42. All reflectivity graphs are plotted after normalization with the help of a Gold mirror at our sample position acting as reference ($R_{905nm}$ = 0.977, $R_{1510nm}$ = 0.966). **Figure 2(c)** compares the measured reflection spectrum with its calculated counterpart. Remarkably, the measured spectrum (blue solid line) exhibits similar features as in the calculated spectrum (red dash-dotted line) for our bridge-type DBR example with 5 polymer/air pairs. Deviations from the model can be understood as a consequence of fabrication imperfections that typically lead to layer-to-layer fluctuations of up to approximately 150 nm (≈ 10 to 15%) around the targeted polymer thickness of about 1.1 µm within the DBR structure, as well as sample-to-sample fluctuations concerning achieved layer thicknesses in the same order of magnitude. Here, a maximum reflectivity of approximately 80% and a stopband width of about 35 nm are obtained at a wavelength about 40 nm shorter than the targeted 935 nm. This shift is attributed to possible and expected shrinkage of the IP DIP material regions in the whole construct after development, with resin contractions specified to be approximately 10-20% according to the product manual. In future studies, the shrinkage can be considered in the design to more accurately meet the target wavelength.

Based on the same underlying print parameters but with an improved (mechanically more rigid, stabilizing) side-pad structure, similar samples containing 5 mirror pairs were additionally fabricated and optically characterized. Again, the experimental data was compared with the numerically calculated spectrum, which predicts higher reflectivity and a wider stopband (red-dashed curve) than observed in experiment due to the considered ideal scenario of a homogeneous Bragg structure without thickness variations. Here, the experimental reflection spectra of two such 5-pair polymer/air-Bragg reflectors, shown in **Fig. 2(d)**, demonstrate higher peak reflectivity (due to better-achieved structure quality) than in (**c**) with a value close to 88% for the 935 nm design, although centered at 905 nm. Furthermore, the very similar spectra of the two printed similar reflectors indicate a reasonably good matching and reproducibility with respect to the achieved layer thicknesses in each stack, and compares well enough with a thickness-variation-based model (shaded). However, due to fabrication imperfections leading for instance to aforementioned experienced thickness fluctuations of typically 150 nm, and possible layer bending/remaining stability issues, there is a discrepancy between theory and experiment.

To showcase the versatile nature of the fabrication process, we have also produced 7λ/4 Bragg mirrors utilizing the aforementioned coaxial design, here with target wavelength 1550 nm (see SEM images in **Fig. 1(d,e)**). The fabricated polymer structures have between 8 and 14 DBR pairs, with one center post and ten side pillars as support structures. The diameter of the center pillar is set to scale one-to-five with the polymer disk diameter (54 µm here).

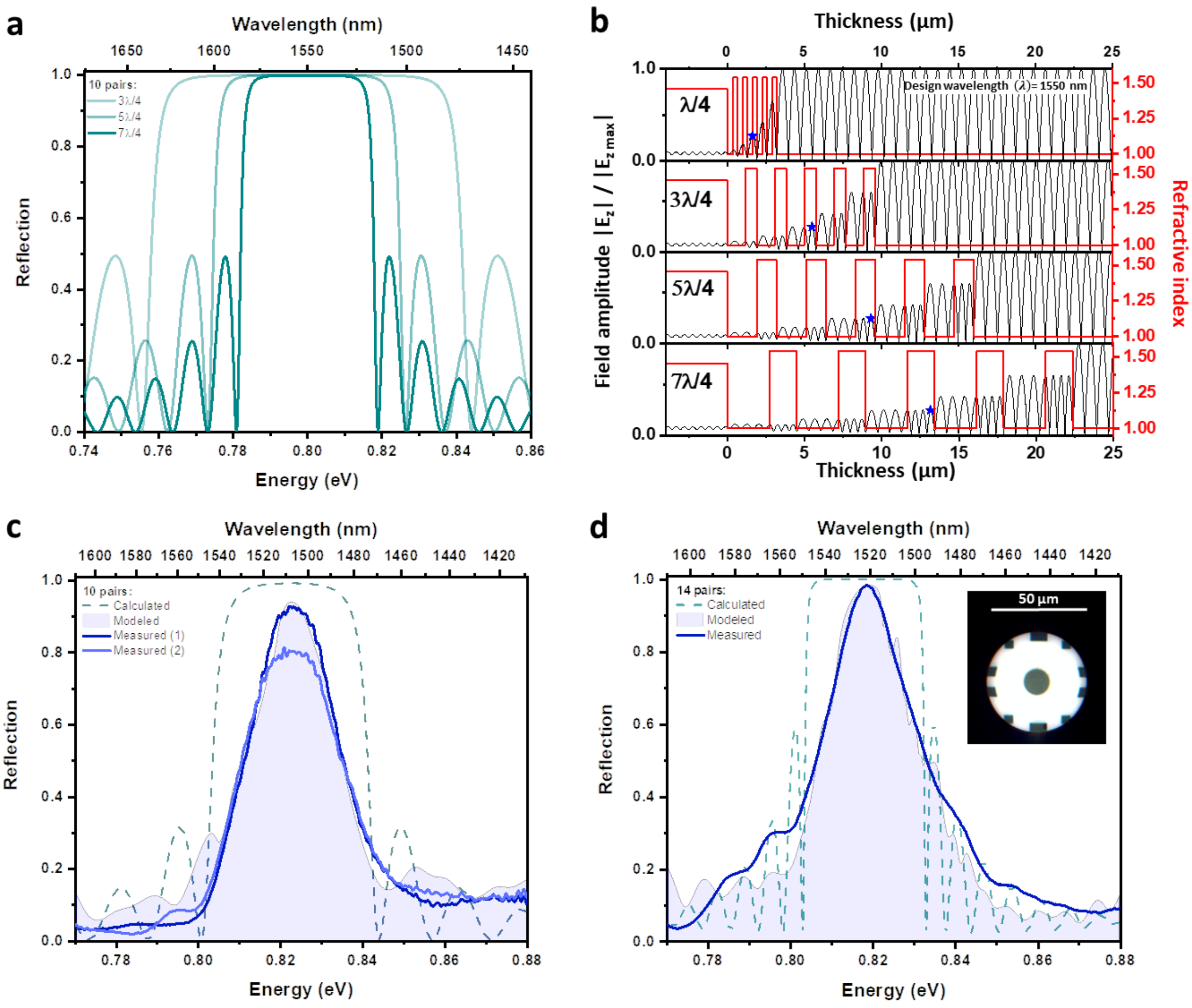

**Figure 3:** Numerically calculated reflectivity (a) and electric-field distribution (b) of 10-layer-pair Bragg configurations for a central wavelength of 1550 nm. (c) Comparison of measured spectra exceeding peak reflectivity of 90% for two coaxial 8-pair air-Braggs on quartz substrate with their theoretical (dashed) as well as modeled (shaded curve) counterpart; the latter taking into account statistical thickness variations. (d) Measured reflectivity for a printed Bragg mirror with 14 pairs and nearly 99% peak

reflectivity, compared to the calculation for an ideal and a model structure. Inset: Close-up white-light microscopy image for the 14-pair structure.

To evaluate the performance of printed coaxial air-Braggs in the C-band wavelength range, similar diagrams as in **Fig. 2** are shown in **Fig. 3**. TMM-calculations indicate the configuration-dependent stopband and field distribution in (**a**) and (**b**), respectively, while experimental results with particularly remarkable peak reflectivity of up to 99% for an all-polymer reflector are shown in **Figs. 3(c) and (d)**. The inset in **Fig. 3(d)** displays an optical microscopy image of a printed coaxial Bragg mirror, in which the support pillars can be seen as dull areas whereas the structures's NIR-wavelength DBR region appears highly reflective even for the visible light.

For reflectors with 8 pairs, the measured reflectivity of the printed Bragg design reaches values in excess of 90% at a central wavelength of about 1510 nm (**Fig. 3(c)**). Here, aforementioned sample-to-sample fluctuations in the achieved polymer and air thicknesses lead to small performance deviations between experimental structures of same type. The 40 nm wavelength blueshift away from the design wavelength is attributed predominantly to the aforementioned resin shrinkage within the development process. The comparison with the numerically calculated reflectivity spectrum of the polymer/air reflector with central wavelength 1510 nm strongly indicates that layer-to-layer thickness fluctuations cause the observed mismatch of measured data with the theoretical stopband behavior – here with up to 400 nm spread between lowest and highest extracted value for air thicknesses as well as polymer thicknesses around 2.8 µm and 1.8 µm, respectively (**Supplementary Fig. S1(a)**). The statistical data in the Supplement not only displays the typical spread in thickness results, but it also leads to the conclusion that the more random these thickness variations occur across the stack, the stronger is the deviation from an ideal DBR stopband spectrum. Alongside the variability in layers' thicknesses, the surface roughness of the printed structures further hinders the overall performance due to possible scattering of light. Moreover, limitations and (spatial) averaging effects in the micro-reflectance measurements can lead to deviations from an ideal focal-spot-independent acquisition of sample back-reflected light.

A striking feature of the coax polymer/air-Bragg reflectors with 14 polymer/air pairs is the fact that they can reach very high peak reflectivity close to 99%, as shown in **Fig. 3(d)** alongside a calculated reflectivity profile with Bragg wavelength of about 1520 nm, especially considering that such printed structures are made up solely of air and polymer layers. A comparison of the theoretical and measured stopbands shows reasonably overlap, suggesting a favorable configuration of polymer and air thicknesses for 1520 nm.

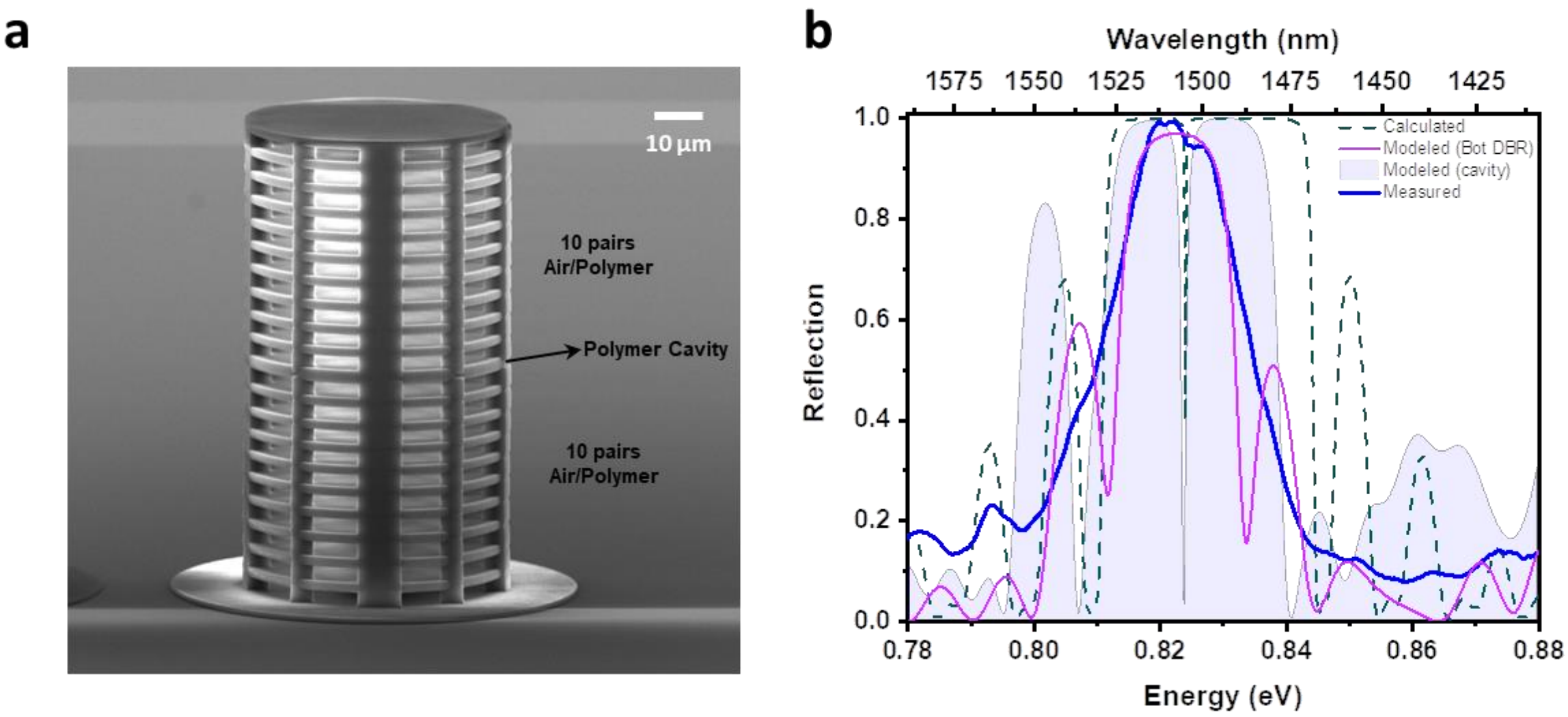

**Figure 4**: (a) SEM of a polymer-air-Bragg reflector-based coaxial microcavity with central polymer full-wavelength cavity design, 3D nanoprinted on quartz substrate for a target wavelength of 1550 nm from two similar 10-pair Bragg reflectors. (b) Top-facet-measured microcavity spectrum (solid dark curve) featuring a peak reflectivity of 99% and a small central cavity dip compared with the theoretical counterpart (dashed, ideal structure), as well as a modeled DBRs-based Fabry-Pérot cavity (shaded curve, statistical thickness variations) and mere bottom DBR (solid light curve).

With our coax design, also a microcavity is implemented with similar top and bottom mirror design and print configuration (**Fig. 4**): The SEM data demonstrates an outstanding fabrication outcome (**a**) with reasonably good matching between experimental (solid) and modeled (shaded) reflectivity features, as well as visible similarities to the ideal case (red dashed curve) (**b**). A strong repeatability is evidenced for similarly printed towers, as **Supplementary Fig. S1(b)** shows.

In summary, we have demonstrated the development and nanofabrication of high-quality polymer-based mirror configurations employing 3D TPL nanoprinting. We showed that single-material DBRs consisting of multi-quarter-wavelength ($7\lambda/4$) alternating layers of air/vacuum and the photoresist polymer printed on dielectric substrates exhibit high reflectivity stopband features in the near-infrared. Both, bridge- and coaxial-type structures showed remarkable peak reflectivity at the Bragg-design wavelength, up to around 99% at 1525 nm and around 85% at 905 nm. In addition, we fabricated a fully microresonator structure in the coaxial resonator design, with a peak reflectivity of 99%. It is expected that such polymer mirrors promise the design and realization of various flexible microcavity configurations that could be engineered and employed for experiments with different types of quantum emitters and sub-micron-scale active media or for environmental-property or mechanical-pressure sensing.

## Funding information

German Research Foundation (DFG: RA2841/5-1, RA2841/12-1—Projektnummer 456700276—, SPP2244 and RE2974/26-1).

## Acknowledgements

Financial support by the Deutsche Forschungsgemeinschaft (DFG) is acknowledged. The authors would like to thank former team member F. Wall for contributing an optimized TMM frame for 2D-materials cavity simulations. CCP would like to thank team members C.-W. Shih and I. Limame for fruitful discussion. CCP and AR-I are grateful to Prof. P. J. Klar for access to the TPL tool, Dr. T. Henning as well as F. Kunze for helpful discussions and insights on 3D nanoprinting and N. Wiegand for technical support in the clean room.

## Conflict of Interest

The authors declare no conflict of interest.

## Author Contributions

**Chirag Chandrakant Palekar**: data curation (lead), formal analysis (lead), software (lead), visualization (lead), resources (lead), investigation (lead), methodology (equal), writing/original draft preparation (lead). **Manan Shah:** methodology (support), investigation (support), resources (support)**. Arash Rahimi-Iman:** conceptualization (lead), methodology (equal), investigation (support), supervision (lead), funding acquisition (equal), validation (equal), writing/review & editing (lead). **Stephan Reitzenstein:** funding acquisition (equal), methodology (supporting), supervision (lead), validation (equal), writing/review & editing (lead).

# Supplement

### S1. Additional considerations of achieved thicknesses

Our investigation revealed that for polymer/air-Bragg mirrors with many layer pairs, the statistics in the practically obtained thicknesses for both polymer and spacer layers causes a persisting discrepancy between numerically calculated values and experimental results owing to fluctuating thickness mismatch distributed across the vertical stack. This can be mainly attributed to imperfections in the print and development process, leading to quite similar reflection performance as well as minor wavelength offsets for achieved Bragg reflectors with layer pairs in excess of 6 pairs (see **Fig. 3(c), (d)**). A statistical evaluation of thickness fluctuations in manually SEM-extracted values is shown in the **Supplementary Fig. S1(a).** For a given print parameter set, the thicker layers are found rather in the lower region of the coaxial reflector tower, whereas thinner layers accumulate at the top. Nonetheless, it cannot be excluded that intermixed mismatch occurs.

A source of general thickness mismatch can also arise from the choice of print parameters used to inscribe certain layer thicknesses into the resin. Furthermore, the shrinking effect for the employed IP-DIP resin can lead to obtainable deviation from the targeted thicknesses.

With an improved design and parameter set used for DBR printing for 1550 nm (cf. **Fig. 3**), fully polymer/air microcavity structures exhibit repeatability of high-quality print results as shown by SEM (see **Fig. S1(b)**).

Additional discrepancy between theory and experiment can further occur due to a structure's possible layer bending/remaining stability issues for free-standing regions, as e.g. indicated for central-pillar supported Bragg towers of extended disk size in the lower part of **Fig. S1(c)**.

In this context it is necessary to note that, we have to realize polymer-layer thicknesses on the level of 1 voxel line scan; air spacers on the scale of 2 polymer-layer thicknesses. For 930-nm samples, the first stable, ultrathin, printable sheet thickness closely matched the 7λ/4 Bragg condition used here, and print speed variations typically allow to thicken and smoothen the hatched and sliced polymer parts to certain extent up to the limits set by the employed TPL machine.

To account for all kinds of fabrication imperfections, further finetuning of print parameters is inevitable and can mitigate some effects, while one can also inversely address certain mismatch by design to counter effects introduced in the production step.

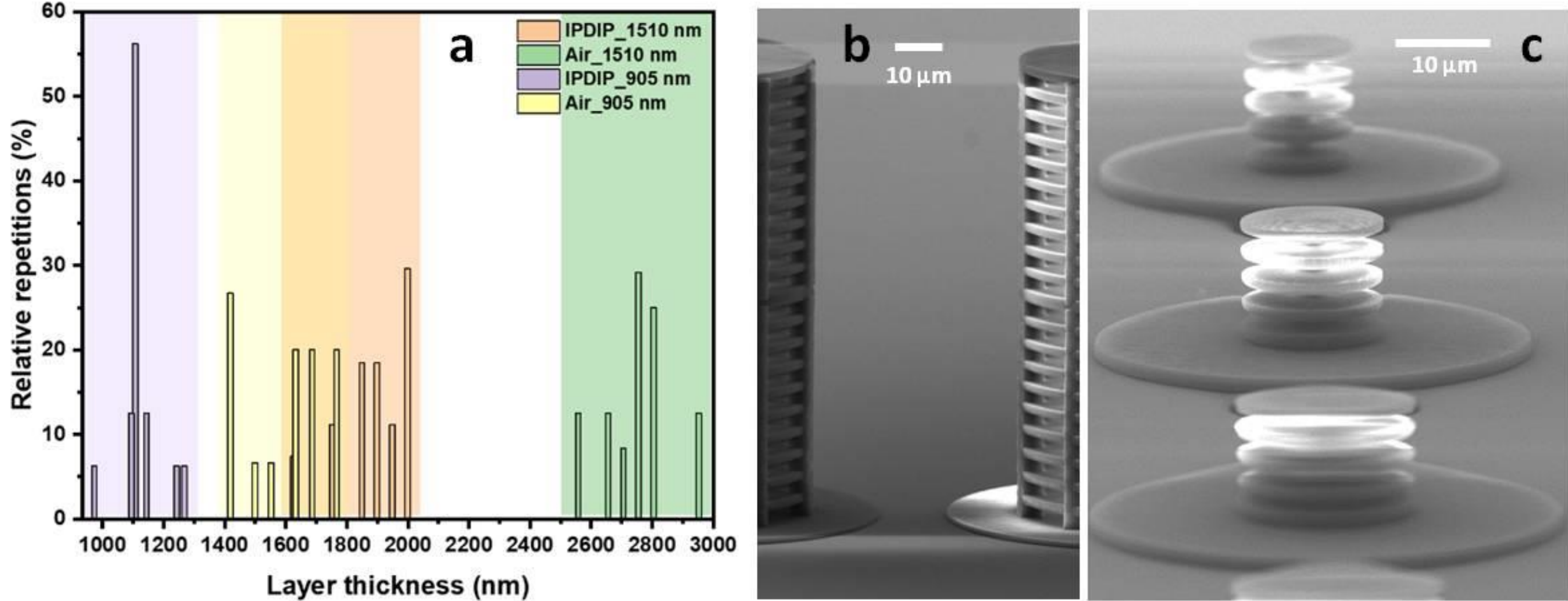

**Figure S1: (a)** The experimental spectra (**Figs. 2 and 3**) for both cases showed central wavelengths at about 905 and 1510 nm, respectively. The histogram reveals the spread in obtained thicknesses for both polymer and spacer layers for both design wavelength cases. Due to the manual read-out from SEM images with a cursor-resolution of about 50 nm, the thickness estimates for the configuration with about 1.1 µm polymer thickness exhibit strongest clustering of values, with the majority of counts at 1.1 µm and relative fluctuations as high as 14%, to lower and higher values. In contrast, for larger layers, the extracted values reveal a larger and more random distribution of thickness values per layer configuration, while the relative fluctuations around the center value go down to as low as about 7%. **(b):** SEM of two similarly printed coaxial polymer/air-Bragg cavities (approx. 90 µm tall) with central 2λ cavity spacer next to each other. **(c)** Disk-size depending stability indication (via SEM) in coax fiber-tip designs for the 1 µm wavelength region towards bending of layers at about 20 µm Bragg disk diameter, when carried by one central few-micron waveguide post, without printed side supports. In **(b)** and **(c)**: Brightness variations due to different conductivity levels among object regions under SEM.